\documentclass[aps,prl,showpacs,amsmath,amsfonts,superscriptaddress,twocolumn]{revtex4}
\usepackage{graphicx}
\usepackage{dcolumn}
\usepackage{bm}
\usepackage{pslatex}
\begin{document}

\title{Microscopic calculation of symmetry projected nuclear level densities}

\author{K. Van~Houcke}
\author{S. M.A.~Rombouts}
\author{K.~Heyde}
\affiliation{Universiteit Gent - UGent, Vakgroep Subatomaire en
  Stralingsfysica \\
  Proeftuinstraat 86, B-9000 Gent, Belgium}
\author{Y. Alhassid}
\affiliation{Center for Theoretical Physics, Sloane Physics Laboratory, Yale
  University\\
  New Haven, Connecticut 06520, U.S.A.}
\date{\today}

\begin{abstract} 
We present a quantum Monte Carlo method with exact projection on parity and
angular momentum that is free of sign-problems for seniority-conserving
nuclear interactions. This technique allows a microscopic 
calculation of angular momentum and parity projected nuclear level densities.
We present results for the $^{55}$Fe, $^{56}$Fe and
$^{57}$Fe isotopes. Signatures of the pairing phase
transition are observed in the angular momentum distribution of the nuclear level density.
\end{abstract}


\maketitle

The level density is a fundamental property of a many-body system as
all thermodynamical quantities can be derived from it.
In nuclear physics, level densities are important because, according to
Fermi's golden rule, they are critical for
estimating nuclear reaction rates.
The simplest approach to estimating the
nuclear level density is to assume a free Fermi gas, leading to the
well-known Bethe formula \cite{Bethe36, BohrMot}. A simple phenomenological
way of incorporating pairing correlations and shell effects is to
back-shift the excitation energy $E_x$ by an amount $\Delta$, resulting in the back-shifted Bethe formula.

The calculation of statistical nuclear reaction rates also requires
knowledge of the angular momentum distribution of the nuclear level density.
An empirical formula for the angular momentum distribution of the level density at a fixed excitation energy $E_x$ assumes uncorrelated and randomly coupled
single-particle spins, and is given by \cite{Bloch54, Ericson60}
\begin{equation}
 \rho_J (E_x) = \rho(E_x) \frac{(2J+1)}{2 \sqrt{2 \pi} \sigma^3}
 e^{-\frac{J(J+1)}{2 \sigma^2}}.
\label{eq:BJ}
\end{equation}
Here $J$ is the angular momentum and $\sigma$ is a spin-cutoff parameter,
which can be used as an energy-dependent fitting parameter. 
The total level density $\rho(E_x)$ in (\ref{eq:BJ}) is often parametrized by a backshifted Bethe formula. The main drawback
of this approach is that it requires a fit for each individual nucleus.

The microscopic calculation of the level density, and in particular its angular momentum distribution beyond the Fermi gas model and the spin-cutoff parametrization
of Eq.(\ref{eq:BJ}) poses a complex
problem. Recently, progress has been made both experimentally \cite{Kalmykov06} and
theoretically \cite{Alhassid06} in gaining more insight into the angular momentum distribution of level densities.
The level density has been calculated microscopically using the shell
model Monte Carlo (SMMC) method \cite{Koonin97, Nakada97}. 
Here, the effect of residual
interactions between the nucleons is taken into account, although restricted
to pairing plus multipole-multipole interactions that are free of the Monte
Carlo sign problem. Angular momentum projection in the SMMC method was
recently carried out for scalar observables by first projecting on the angular
momentum component $J_z$ \cite{Alhassid06}. Angular momentum projection
introduces a new sign problem for $J \neq 0$ even when for a good-sign
interaction. At intermediate and high temperatures, this sign problem is
sufficiently moderate to allow the reliable derivation of angular momentum
distributions. However, at low temperatures this sign problem becomes more
severe.

In this Letter, we present a new breakthrough that avoids
sign problems at any value of angular momentum and/or particle number
and that allows a microscopic test for the validity of the Fermi gas and
spin-cutoff model for nuclei that exhibit strong pairing correlations.
We discuss a quantum Monte Carlo (QMC) method to solve a general isovector $J=0$ pairing
model and solve the angular momentum projection problem for
seniority conserving models.
The pairing model serves as a benchmark to identify signatures of pairing correlations in the
nuclear level density. In a finite nucleus, the number of paired particles is relatively small,
and the identification of such signatures of the nuclear superfluid
transition is difficult. Including higher order interaction multipoles,
would also introduce a sign problem in our approach. Nonetheless, the projection is direct, 
and our QMC method allows us to treat the very large
model spaces, which are needed for the calculation of level densities at higher excitation
energies, and, particularly, for the study of the parity and
angular momentum dependence of level densities (even at low excitation energy).

We start from a general isovector $J=0$ pairing model, here written
in the quasi-spin formalism \cite{Dean03, Brink06}
\begin{eqnarray}
  \hat{H} & = & \sum_j 2\varepsilon_j \hat{S}^0_j - \sum_{jj'} g_{jj'} \hat{S}^+_j \hat{S}^-_{j'},
  \label{eq:pairham}  \\
\hat{S}^0_j & = &   \frac{1}{2} \sum_{m>0} \big( \hat{a}^{\dag}_{jm}\hat{a}_{jm} +
\hat{a}^{\dag}_{j\bar{m}}\hat{a}_{j\bar{m}} - 1  \big), \nonumber \\
\hat{S}^{+}_j & = & \sum_{m>0} \hat{a}^{\dag}_{jm} \hat{a}^{\dag}_{j\bar{m}},
\quad \hat{S}^{-}_j  =  \sum_{m>0} \hat{a}_{j\bar{m}} \hat{a}_{jm},
\end{eqnarray}
where the operator $\hat{a}^{\dag}_{jm}$ creates a particle in the spherical mean-field
single-particle state $|jm\rangle$ with energy $\varepsilon_j$, and $|j
\bar{m}\rangle$ is the time-reverse conjugate state of $|jm\rangle$.
The operators $\hat{S}^0_j, \hat{S}^+_j, \hat{S}^-_j$ close an SU(2) algebra and are known as quasi-spin operators.
The Hamiltonian (\ref{eq:pairham}) is exactly
solvable for a constant pairing strength.
The QMC method discussed here allows us to solve the general case when all terms in (\ref{eq:pairham}) are attractive, i.e., $g_{jj'}>0$.

We formulate the pairing problem in the quasi-spin formalism to address the
problem of the angular momentum projection.  
The quasi-spin projection $S^0_j$
determines the total number of particles $N_j =
2S^0_j+ \Omega_j$ in level $j$ ($\Omega_j = j+1/2$ is the pair degeneracy of
orbital $j$), while the quasi-spin quantum number $S_j$ determines the
seniority quantum number $\nu_j = \Omega_j - 2S_j$, i.e., the number of unpaired particles in level $j$.
Within this quasi-spin (or seniority) scheme, angular momentum remains a good
quantum number. Given a set of quasi-spin quantum numbers $S_j$, one can
determine the degeneracy for a given value of the total angular momentum $J$.
Formally, the problem is equivalent to a chain of spins. The pairing
interaction can flip spins such that the total ``magnetization'' $\sum_j
S^0_j$ remains constant (particle number conservation). In addition,
each quasi-spin quantum number $S_j$ can take values between $0$ and $\Omega_j/2$.

There exist a number of very efficient QMC approaches to simulate spin
chains \cite{Kawashima04}. However, in our quasi-spin model, the value of $\sum_j
S^0_j$ is fixed, while the quasi-spin values $S_j$ can change. Recently, we
developed a non-local loop update QMC scheme that is capable of sampling spin
models at constant magnetization. In our case this means sampling the canonical ensemble 
\cite{Rombouts05}. Assuming the Hamiltonian $\hat H = \hat H_0 -\hat V$ consists of two non-commuting parts
$\hat{H}_0$ and $\hat{V}$, our scheme is based on a perturbative expansion of the partition function at inverse temperature $\beta$
\begin{eqnarray}
{\rm{Tr}} \big( e^{-\beta \hat{H}} \big)
     & = & \sum_{m=0}^{\infty}
    \int_{0}^{\beta} d \tau_m \int_0^{\tau_m} d \tau_{m-1} \cdots  \int_0^{\tau_2} d \tau_1
     \nonumber\\
    & &  {\rm{Tr}} \big[\hat{V}(\tau_1)  \hat{V}(\tau_2)
 \cdots   \hat{V}(\tau_m)
       e^{-\beta \hat{H}_0}\big],
       \label{eq:decompos}
\end{eqnarray}
with $\hat{V}(\tau) = $exp$(-\tau\hat{H}_0) \hat{V} $exp$(\tau\hat{H}_0)$. Here we choose
\begin{equation}
\hat{H}_0  =  \sum_j 2\varepsilon_j \hat{S}^0_j - \sum_{j} g_{jj} \hat{S}^+_j \hat{S}^-_{j}\;;\;\;\;\;\; \hat V = \sum_{j \neq j'} g_{jj'} \hat{S}^+_j \hat{S}^-_{j'} \;.
\end{equation}
The basic idea of the QMC method is to insert a so-called worm operator $\hat{A}$ in the
partition function, obtaining an extended partition function ${\rm{Tr}}
\big(\hat{A}e^{-\beta\hat{H}} \big)$.
By propagating this worm operator through imaginary
time according to the rules explained in Ref.~\cite{Rombouts05},
one generates configurations that are
distributed according to their weight in
the canonical partition function ${\rm{Tr}}_N
\big( e^{-\beta\hat{H}}\big)$ at fixed particle number $N$.
Here, the worm operator is chosen to be $ C + \sum_{ij} S^+_i S^-_j$, with $C$ a
constant. Such a worm operator
allows for the sampling of all configurations that correspond to a fixed set
of quasi-spin quantum numbers $S_j$, without changing the value of $\sum_j S^0_j$.
However, to be ergodic,
the worm operator must also generate configurations with varying seniority
quantum numbers, and therefore we add a worm operator that
can change the values of the quasi-spins $S_i$, $S_j$ and their projections $S^0_i$, $S^0_j$ for two levels $i$ and
$j$ such that
$(S^0_i+S^0_j)$ remains the same (particle-number conservation).
The complete quasi-spin
phase space is now sampled by propagating a seniority conserving
worm operator and an additional seniority non-conserving worm operator.
Additional details on the worm operator moves can be found in Ref.~\cite{Rombouts05}.

The moves of the worm operator can be constrained in such a way that the
angular momentum $J$ of the configuration does not change. Choosing the initial
configuration with well defined angular momentum quantum number $J$ enables us to calculate thermal
averages for that fixed value of $J$. Parity
is also a good quantum number in the quasi-spin basis, so that parity projection can be accomplished in a similar way.

We have used the QMC method outlined above to study the angular momentum and parity
distribution of nuclear level densities in the presence of pairing correlations.
We focused on the iron isotopes $^{55}$Fe,  $^{56}$Fe and $^{57}$Fe,
within the complete $0f+1p+0g_{9/2}$ model space. To study truncation effects,
we also consider an extended model space $1s+0d+0f+1p+0g_{9/2}+2s+1d$.
The inverse temperature $\beta$ ranges from $0$ to $2.5$ MeV$^{-1}$.
For the mean-field potential, we used a Woods-Saxon potential with the
parametrization of Ref.~\cite{Perey72}.
The single particle energies $\epsilon_{j}$ were obtained by diagonalizing this potential in a harmonic oscillator basis.
For simplicity, we used a constant pairing strength (although our QMC method is not
restricted to this case), determined to reproduce the experimental gap parameter
$-1/2 (\mathcal{B}(N-1,Z) - 2 \mathcal{B}(N,Z) + \mathcal{B}(N+1, Z))$ for $Z=26$ protons and $N=30$
neutrons, with $\mathcal{B}$ denoting the binding energy \cite{BohrMot}. 
For the larger valence space, the value of pairing strength $g$ is
renormalized in such a way that the gap parameter remained fixed.
Using the temperatures for which the heat capacity is maximal, we estimate our calculations to be free of truncation effects for excitation energies up to $E_x
\lesssim 20$ MeV and $E_x\lesssim 50$ MeV for the
$pf+g_{9/2}$ and
$sd+pf+g_{9/2}+sd$ valence spaces, respectively.
Total and angular momentum/parity-projected level densities are calculated as usual in the saddle point approximation.

\begin{figure}[ht]
\begin{center}
\includegraphics[angle=0, width=8.5cm] {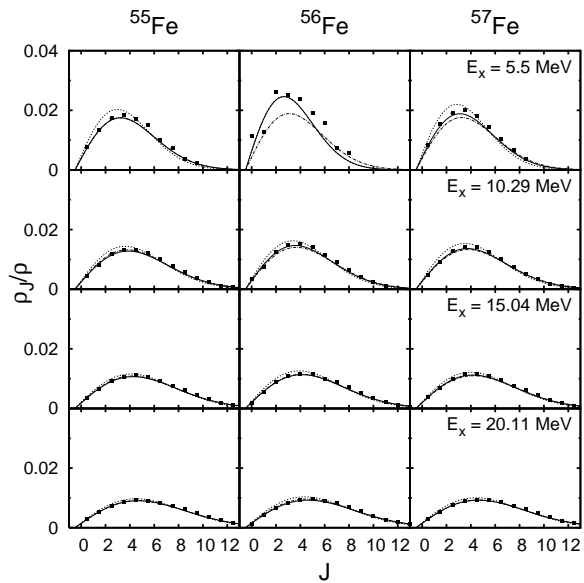}
\caption{The angular momentum distribution of the level density at a given excitation
  energy $E_x$ for the iron isotopes $^{55}$Fe, $^{56}$Fe and $^{57}$Fe. Solid
  squares arethe Monte Carlo results in the $pf+g_{9/2}$ valence space. The
  solid lines are fits to the spin-cutoff model Eq.~(\ref{eq:BJ}). Also shown
  are distributions given by the spin-cutoff model with $\sigma^2$ calculated
  from Eq.~(\ref{eq:moi}) using the rigid body moment of inertia (dotted-dashed lines), and from
  Eq.~(\ref{eq:j2}) (dotted lines).
   \label{fig:spindistr3Fe}}
\end{center}
\end{figure}

Figure \ref{fig:spindistr3Fe} shows the
angular momentum distribution of the projected level density $\rho_J$  at four
different excitation energies $E_x$ (we do not include not the magnetic
degeneracy in $\rho_J$, 
i.e. $\rho(E_x) = \sum_J (2J+1) \rho_J(E_x)$). The solid squares are the QMC results (the statistical error are much smaller than the size of the squares), while
the solid lines are the spin-cutoff model
Eq.~(\ref{eq:BJ}) fitted to the QMC data with $\sigma^2$ as a fit
parameter. For all three iron isotopes, the angular momentum distribution
becomes broader with increasing excitation energy. When this energy is
sufficiently high ($E_x \geq 10$ MeV), the QMC data are well described by the
spin-cutoff model. For excitation energies $\leq 10$ MeV, some deviations are
observed. In particular, a staggering effect (in $J$) is found for the even-even $^{56}$Fe isotope but not in the odd-even Fe isotopes.
This effect was also reported in Ref.~\cite{Alhassid06}, where the
interaction included higher multipoles, and was recognized as a
signature of pairing correlations.
Here, we observe a strong suppression of the $J=1$ level density ($\rho_1 \approx
\rho_0$), which is much smaller than the $J=2$ level density. For higher
angular momenta, we found only a very small staggering in the angular momentum
dependence. This can be understood from the fact that the pairing interaction
scatters only $J=0$ nucleon pairs.

In a semiclassical approach, the spin-cutoff parameter is given by
\begin{equation}
  \sigma^2 = \frac{1}{3} \langle \hat{J}^2  \rangle,
\label{eq:j2}
\end{equation}
with $\langle \ldots \rangle$ the thermal average at temperature $T$ \cite{Alhassid05}. It is then common to define
an effective moment of inertia from
\begin{equation}
  I = \frac{\hbar^2}{T} \sigma^2.
\label{eq:moi}
\end{equation}
The spin-cutoff parameter is often determined by substituting the
rigid-body moment of inertia, $I = 2 m A (r_0 A^{1/3})^2 /5$ in Eq.~(\ref{eq:moi}). (Here, $r_0$ is the
nuclear radius parameter, $A$ is the mass number and $m$ is the nucleon
mass). The spin-cutoff model (\ref{eq:BJ}) with $\sigma^2$
calculated from the rigid body moment of inertia (dotted-dashed lines in Fig.~\ref{fig:spindistr3Fe})
essentially coincides for $E_x \geq 10$ MeV with the distribution fitted to the Monte
Carlo data (solid lines in Fig.~\ref{fig:spindistr3Fe}).
For $^{56}$Fe, however, for $E_x<10$ MeV, the rigid-body moment of inertia
predicts a broader angular momentum distribution than the one described by the fit. This
indicates a reduction of the effective moment of inertia because of
pairing correlations, and is 
a signature of nuclear superfluidity.
We also determined $\sigma^2$ from the thermal average $\langle \hat{J}^2
\rangle$, calculated directly in the QMC simulation.
The angular momentum distributions using these spin-cutoff parameter are
shown in Fig.~\ref{fig:spindistr3Fe} by the dotted lines. 
These distributions are slightly more peaked than the fitted distributions
(solid lines), especially at low excitation energies ($E_x \leq 10$ MeV). For
$^{56}$Fe at $E_x = 5.5$ MeV, this curve coincides with the fit.

\begin{figure}[ht]
\begin{center}
\includegraphics[angle=0, width=8.5cm] {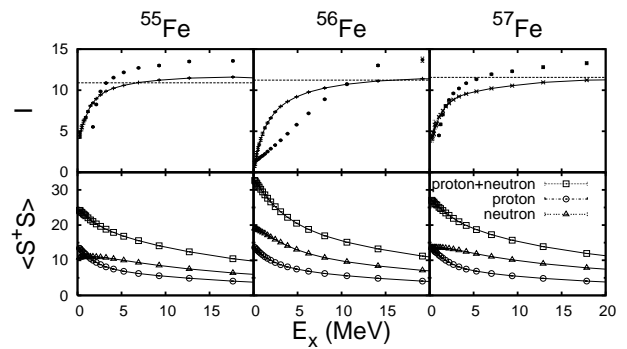}
\caption{Top panels: the moment of inertia $I$ (in units $\hbar^2$) for $^{55}$Fe (left),
  $^{56}$Fe (middle) and $^{57}$Fe (right). The horizontal dashed lines are the rigid
  body moments of inertia, the solid lines are moments of inertia calculated
  from Eq.~(\ref{eq:j2}), and the solid circles are from Eq.~(\ref{eq:moi}) with
  $\sigma^2$ determined via $\rho_{0}/\rho$ (in the even-even case) and
  $\rho_{1/2}/\rho$ (in the odd-even case). 
  Bottom panels: the pair correlation energy $\langle S^\dagger S\rangle$.
   \label{fig:J23Fe}}
\end{center}
\end{figure}

The top panels of Fig.~\ref{fig:J23Fe} shows the effective moment of
inertia as a function of excitation energy. At high excitation energy, the
moment of inertia calculated from $\langle \hat{J}^2
\rangle$ is very close to its rigid body value, the latter indicated with the
horizontal dashed lines. We also calculated the spin-cutoff parameter $\sigma^2$ directly from the ratio
$\rho_{0(1/2)}/\rho$ in the even (odd) case. The corresponding moments of
inertia are also shown in the top panels of Fig.~\ref{fig:J23Fe} (solid
circles). These values seem to differ from our previous estimates of the moment of inertia,
but they do not lead to significant differences in the angular momentum 
distributions, except for $^{56}$Fe for $E_x<10$ MeV. Here, the moment of inertia is strongly suppressed
compared to $^{55}$Fe and $^{57}$Fe, due to $J=0$ pairing. 

The bottom panels of Fig.~\ref{fig:J23Fe} show the pair
correlation energy $\langle S^+S^- \rangle = \langle \sum_{j, j'} S^+_j
S^-_{j'} \rangle$ as a function of excitation energy.
It is seen that the pair correlation energy is strongly suppressed
with increasing excitation energy.
The suppressed neutron correlation energy of $^{55}$Fe and $^{57}$Fe is seen to increase first
slightly with increased excitation energy (or temperature). This results from a blocking effect of the unpaired neutron at low temperatures.

\begin{figure}[ht]
\begin{center}
\includegraphics[angle=0, width=8.5cm] {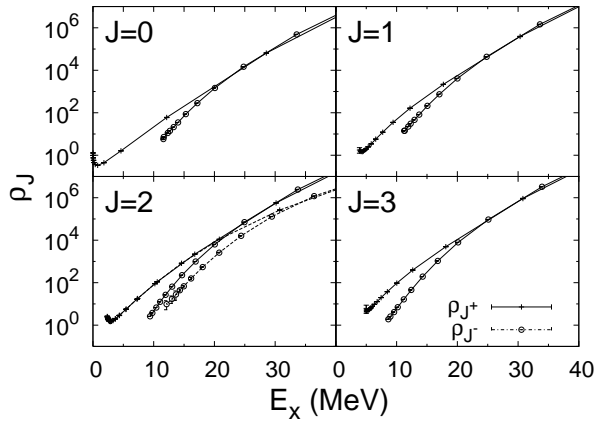}
\caption{The angular momentum and parity-projected level density of $^{56}$Fe
  for the four lowest angular momentum values. A strong parity dependence is found for excitation
  energies up to $\sim 20$ MeV. The valence model space is $sd+pf+g_{9/2}+sd$. For
  $J=2$, we also show results for the smaller $pf+g_{9/2}$ valence space (dashed lines).
   \label{fig:spinpar}}
\end{center}
\end{figure}

In Fig.~\ref{fig:spinpar} we show the parity-projected level density for $J=0,1,2,3$ as a function of excitation energy
using the $sd+fpg+sd$ valence space. For energies
below $\sim 20$ MeV, the angular momentum projected level densities show a strong parity
dependence. Recently, angular momentum and parity-projected level densities were
determined experimentally \cite{Kalmykov06}. However, no parity dependence was
resolved for level densities with specific values of $J$. Our calculations show that
the angular momentum projected level density displays a strong parity dependence.
For $J=2$, we also show the level density within the smaller $pf+g_{9/2}$
valence space (dashed lines). Below $\sim 20$ MeV, the even-parity level
density is in good agreement with the results found in the $sd+fpg+sd$ model
space. However, the inclusion of the $sd$ shells, below and above the
$pf+g_{9/2}$ shell, significantly enhances the odd-parity level density at low
excitations. We found a similar effect for the total odd-parity level density.
This results from the increased fraction of single-particle levels with positive parity.

\begin{figure}[ht]
\begin{center}
\includegraphics[angle=0, width=8.5cm] {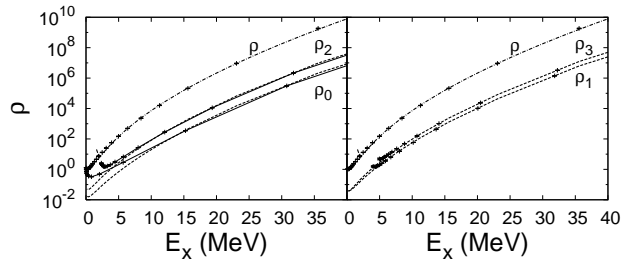}
\caption{The QMC total level density for $^{56}$Fe, well described by the backshifted Bethe formula (dotted-dashed line), together with the projected densities $J=0$,
  $J=2$ (left panel), and $J=1$, $J=3$ (right panel). 
The solid lines connect the $J$-projected Monte Carlo results, while the dashed lines show the
projected level density from the spin-cutoff model Eq.~(\ref{eq:BJ}) using
the rigid body values of $\sigma^2$.
   \label{fig:nld54Fe}}
\end{center}
\end{figure}

In the left (right) panel of Fig.~\ref{fig:nld54Fe} we show the QMC total level density,
as well as the angular momentum projected level densities $J=0$ ($J=1$)
and $J=2$ ($J=3$) for $^{56}$Fe. The QMC total level density is well described
by the backshifted Bethe formula (fitted for 
$4.5$ MeV $< E_x
<$ $40$ MeV) with Fermi gas parameter $a = 5.741 \pm 0.034$ MeV$^{-1}$ and backshift
$\Delta = 1.591 \pm 0.057$ MeV (see dotted-dashed line in
Fig.~\ref{fig:nld54Fe}). These are similar to the SMMC level density
parameters $a = 5.780 \pm 0.055$ MeV$^{-1}$ and $\Delta = 1.560 \pm 0.161$ MeV
of Ref.~\cite{Nakada97}, which were found to be in good
agreement with the experimental level density. The present results indicate that the 
backshifted Bethe formula with temperature-independent parameter $a$ works 
well up to higher energies \cite{Alhassid03}.
The dashed lines of Fig.~\ref{fig:nld54Fe} are the angular momentum projected
level densities, calculated from the total level density through
Eq.~(\ref{eq:BJ}) with rigid body $\sigma^2$. At high excitation energy, these
densities coincide with the QMC densities. However, for excitation energies
below $\sim 10$ MeV, the QMC $J=0$ level density deviates from the level
density predicted by the spin-cutoff model with rigid body $\sigma^2$. This is
a signature of the pairing phase
transition in the level density. This signature is also visible in the $J=2$
level density for energies below $\sim 8$ MeV. For the odd $J$ values, however, all
QMC data are well described by the rigid body model.

In conclusion, we have used a 
quantum Monte Carlo method to
calculate angular momentum and parity-projected level densities for a nuclear
pairing model. This method allows us to calculate level densities in the very
large model spaces that are necessary to avoid truncation effects.
We have found that pairing correlations affect the angular momentum distribution of the level
density at low excitation energies, thereby revealing signatures of the
pairing phase transition in nuclei.

We thank N. Jachowicz and K. Langanke for interesting suggestions and
discussions. We acknowledge the financial support of the Fund for Scientific
Research - Flanders (Belgium) and IUAP, and of the U.S. DOE grant No.\
DE-FG-0291-ER-40608.

\end{document}